\documentstyle[aps,epsf,floats,twocolumn]{revtex}

\tighten
\begin{document}
\draft

\title{Impurity scattering and localization in $d$-wave superconductors}

\author{M. Franz, C. Kallin and A. J. Berlinsky}
\address{Department of Physics and Astronomy and Brockhouse Institute for 
Materials Research,\\  McMaster University,  Hamilton, Ontario, L8S 4M1 
Canada \medskip\\
\parbox{14cm}{\rm
Strong evidence is presented for the localization of
low energy quasiparticle states in disordered $d$-wave superconductors. 
Within the framework of the
Bogoliubov-de Gennes (BdG) theory applied to the extended Hubbard model
with a finite concentration of non-magnetic impurities,
we carry out  a fully self-consistent numerical diagonalization 
of the BdG equations on finite clusters containing up to $50\times 50$ sites. 
Localized states are identified 
by probing their sensitivity to the boundary conditions and by analyzing
the finite size dependence of inverse participation ratios.\\
}}

\maketitle

\narrowtext
In conventional $s$-wave superconductors nonmagnetic impurities have little
effect on the superfluid density and the transition temperature, 
which may be understood from  Anderson's theorem\cite{at}. The 
situation is dramatically different in the high temperature cuprate 
superconductors where  nonmagnetic impurities exhibit a strong pair breaking
effect \cite{ueda}. 
This is interpreted as a manifestation of unconventional, most likely 
$d$-wave pairing symmetry. Understanding the role that impurities 
play in such superconductors is crucial for the interpretation of experimental
data. A good example of this is the interpretation of experiments measuring
the temperature dependent penatration depth \cite{pend,hirschfeld}
in YBa$_2$Cu$_3$O$_{7-x}$. 
The behavior of various quantities in the presence of disorder also
serves as an important test for theoretical models of microscopic pairing 
interaction.

In the present paper we concentrate on the possibility of localization of the
low-energy quasiparticle states in $d$-wave superconductors with line nodes
in the gap on the fermi surface. We wish to address the question of whether
these low lying states are strongly localized with a relatively short
localization length $\xi_L$ or whether they are essentially extended. This is
a question  of significant importance for many experiments since the
transport and thermodynamic properties are largely determined by these low
energy excitations.

The problem of localization in $d$-wave superconductors was first considered by
Lee \cite{lee} who, appealing to arguments from the scaling theory
of localization, found that in the limit of unitary ({\it i.e.} strong) 
scatterers the
quasiparticle states are strongly localized below the mobility gap ${\gamma_0}$, even
if the impurity concentration is sufficiently 
small that the normal state wavefunctions
are essentially extended. For moderate concentrations of strong scatterers, 
${\gamma_0}$ is a reasonable fraction of the maximum gap which allows
for the possibility of experimental confirmation. One consequence of such a 
scenario is the prediction of a universal limit of conductivity
$\sigma(\omega\to 0) \simeq (e^2/2\pi\hbar)\xi_0/a$,  independent of the 
scattering rate $\tau^{-1}$ ($\xi_0$ is a coherence length and $a$ is the 
lattice constant). Infrared reflectance data on ion-irradiated 
YBa$_2$Cu$_3$O$_{7-x}$ have been interpreted using this picture \cite{basov}.

More recently Balatsky and Salkola \cite{balatsky} presented 
results that contradict this scenario. Their argument is based on a physical 
picture of a single impurity wavefunction which, according to an earlier
calculation within the self-consistent $T$-matrix approximation\cite{balatsky95},
is highly anisotropic with slowly decaying ($\sim 1/r$) tails along the
(11) and ($1\bar{1}$) diagonals. 
Overlaps between these tails lead to strong interactions between
the quasiparticle states on distant impurities, which then form a network
of extended impurity states capable of carrying current. This network
percolates across the entire system and inhibits the localization by disorder
at the lowest energies, creating an ``inverse mobility gap'' $\gamma_c<\gamma_0$.

Both theories described above resort to approximations when treating the 
scattering from impurities and they ignore the gap relaxation near the
impurity sites resulting from their pair breaking property. Since the arguments
for and against localization are quite subtle it may well be that the above
mentioned details are important, in which case perhaps the best way to 
address the problem is through  numerical calculations. Hatsugai and Lee
\cite{hatsugai} studied localization numerically in a simpler but related
model of Dirac fermions on the lattice, whose excitation spectrum is
similar to that of a $d$-wave superconductor. By examining the
sensitivity of the wavefunction to boundary conditions they concluded
that the low energy states are indeed strongly localized. This would seem to be
in qualitative agreement with the theory of Lee\cite{lee}, however since
the system possessed no off-diagonal long range order, the quantitative
predictions of this theory could not be verified. While  this work was 
strongly suggestive,
without treating a system with superconducting order one cannot establish
the existence of  
localization in a $d$-wave superconductor. A convenient framework for studying
the effect of impurities within a simple lattice model of $d$-wave 
superconductivity was outlined by Xiang and Wheatley \cite{xiang}.
The same model was later used to study localization induced by {\em weak} 
disorder, modeled by a white noise random component added to the chemical 
potential \cite{xiang1}. This corresponds to the scattering in the Born limit
where the localization effect had been predicted to be negligible \cite{lee}.
The numerical work indeed confirmed that the localization length in the
superconducting state remains comparable to its normal-state 
value ({\it i.e.}, very large) down to the lowest energies.

In the present work we address the case of a dilute
 density of {\em strong} (nearly
unitary) scatterers. It is this limit that 
has precipitated controversy \cite{lee,balatsky}, and that is most relevant 
experimentally ({\it e.g.}, for Zn doped YBa$_2$Cu$_3$O$_{6-\delta}$). 
In order to confront the question of localization in such systems
we solve a simple tight binding model with an on-site
repulsion and nearest neighbor attraction which give rise to superconductivity
in the $d$-wave channel, on finite clusters
using the selfconsistent Bogoliubov-de Gennes (BdG) technique. The main advantage
of such an approach is that the 
impurity scattering is treated {\em exactly} and that
it is {\em directly } relevant to $d$-wave superconductors. Our principal
result is that quasiparticle states are strongly localized at low energies 
below a 
mobility gap ${\gamma_0}$. The value of ${\gamma_0}$, as well as the localization length 
$\xi_L$, are in a good agreement with the theory of Lee\cite{lee}.

The Hamiltonian we consider has been used previously to study the vortex
structure \cite{soininen,wang} and impurities \cite{xiang,xiang1} in a $d$-wave
superconductor:
\begin{eqnarray}
H=&-&t\sum_{\langle ij \rangle\sigma} c^\dagger_{i\sigma}c_{j\sigma}   
  -\mu\sum_{i\sigma}n_{i\sigma}
  +\sum_{i\sigma}V^{imp}_i n_{i\sigma}  \nonumber \\
  &+&V_0\sum_i n_{i\uparrow}n_{i\downarrow}
  +{V_1\over 2}\sum_{\langle ij \rangle\sigma\sigma'}n_{i\sigma}n_{j\sigma'}
\label{hub}
\end{eqnarray}
Here $\langle ij \rangle$ stands for nearest neighbor pairs, and 
the notation is otherwise standard. The last term models strongly repulsive 
impurities: $V_i^{imp}=V^{imp}>0$ at randomly chosen sites with  density
$n_{imp}$ and $V_i^{imp}=0$ on all other sites.
If one takes $V_0>0$ and $V_1<0$  this model gives rise to
pairing in the $d$-wave channel. In mean field
theory (\ref{hub}) can be solved by defining the pairing amplitudes 
\begin{equation}
\Delta_0({\bf r}_i) = V_0\langle c_{i\uparrow}c_{i\downarrow} \rangle, \ \ \
\Delta_\delta({\bf r}_i) = 
V_1\langle c_{i+\delta\uparrow}c_{i\downarrow} \rangle, \nonumber 
\end{equation}
where $\delta=\pm\hat{x}, \pm\hat{y}$ are nearest neighbor vectors for a square 
lattice. 
The resulting mean field Hamiltonian is then diagonalized using the Bogoliubov
transformation\cite{degennes}
\begin{eqnarray}
c_{i\uparrow}&=&\sum_n[\gamma_{n\uparrow}u_n({\bf r}_i)
                      -\gamma_{n\downarrow}^\dagger v_n^*({\bf r}_i)], \nonumber \\
c_{i\downarrow}&=&\sum_n[\gamma_{n\downarrow}u_n({\bf r}_i)
                      +\gamma_{n\uparrow}^\dagger v_n^*({\bf r}_i)]. \nonumber 
\end{eqnarray}
to the quasiparticle operators $\gamma_{n\sigma}$. Within mean field theory 
the Hamitonian  (\ref{hub})
is diagonalized when $u_n$ and $v_n$ satisfy the BdG equations 
\cite{degennes,soininen}
\begin{equation}
\left( \begin{array}{cc}
\hat{\xi}      & \hat{\Delta} \\
\hat{\Delta}^*  & -\hat{\xi}^*
\end{array} \right)
\left( \begin{array}{c}
u_n \\ v_n
\end{array} \right)
=E_n 
\left( \begin{array}{c}
u_n \\ v_n
\end{array} \right),
\label{BdG}
\end{equation}
where
\begin{eqnarray}
\hat\xi u_n({\bf r}_i) &=& -t\sum_\delta u_n({\bf r}_i+\delta) 
+(V_i^{imp}-\mu)u_n({\bf r}_i), \nonumber \\
\hat\Delta v_n({\bf r}_i)&=&\Delta_0({\bf r}_i) 
+\sum_\delta \Delta_\delta({\bf r}_i)v_n({\bf r}_i+\delta),
\end{eqnarray}
 subject to the constraints of self-consistency
\begin{eqnarray}
\Delta_0({\bf r})   &=& V_0\sum_n u_n({\bf r})v_n^*({\bf r}) \tanh(E_n/2k_BT), \nonumber \\
\Delta_{\bf \delta}({\bf r}) &=& {V_1\over 2} \sum_n\left[u_n({\bf r}+{\bf \delta})v_n^*({\bf r})
                                 +u_n({\bf r})v_n^*({\bf r}+{\bf \delta})\right]    \nonumber \\
                && \ \ \ \ \ \ \ \ \ \ \ \times \tanh(E_n/2k_BT),
\label{self}
\end{eqnarray}
where the summation is over positive eigenvalues $E_n$ only.
\begin{figure}
\epsfxsize=7cm
\epsfysize=14cm
\epsffile{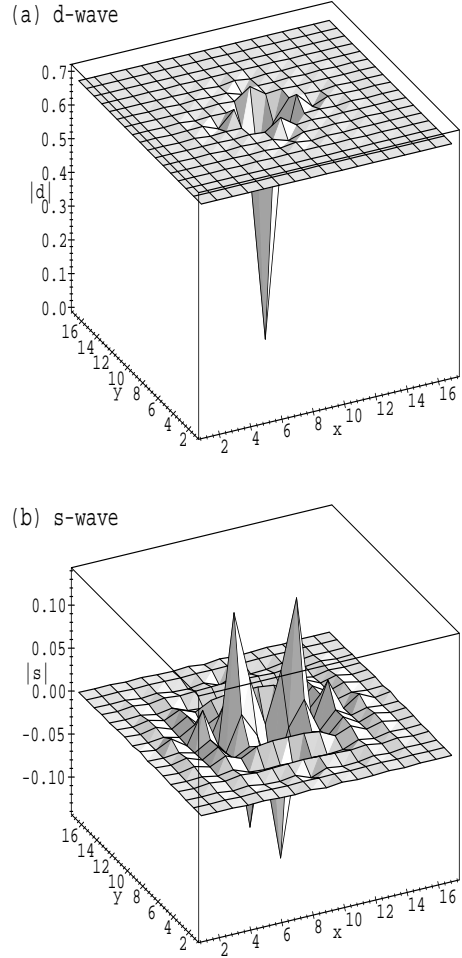}
\caption[]{Amplitudes of a) $d$-wave and b) $s$-wave order parameters
in the vicinity of a single impurity for a $17\times 17$ system. Note the 
different scales for $d$ and $s$. \label{ds}}
\end{figure}

For a system of linear size $L$, solving the system of equations (\ref{BdG}) 
with periodic boundary conditions
requires diagonalizing a $2L^2\times 2L^2$ matrix. For a suitably chosen
initial order parameter distribution we solve the system (\ref{BdG}) using the 
standard  LAPACK diagonalization routine. We than compute new gap functions
from Eqs.\ (\ref{self}) and iterate until the desired convergence is achieved.
Typically 7-8 iterations are necessary to establish five-digit 
convergence in the free energy and the average gap. Periodically we perform 
longer runs of 25-30 iterations to confirm the accuracy of the solutions. No 
significant deviations are found between short and longer runs.

We have solved the BdG equations (\ref{BdG}) at $T=0$
for the following set of model 
parameters: $\mu=-t$, corresponding to the band filling factor $n\simeq 0.68$
($n=1$ is a half-filled band); $V_0=3t$, and $V_1=-4.5t$. In the absence of 
disorder Eqs.\  (\ref{BdG}) are easily solved analytically by appealing to 
translational invariance. One obtains the usual BCS type excitation spectrum
$E_{\bf k}=(\epsilon_{\bf k}^2+|\Delta_{\bf k}|^2)^{1/2}$ with 
$\epsilon_{\bf k}=-2t(\cos k_x + \cos k_y)-\mu$ and 
$\Delta_{\bf k}=2\Delta_d(\cos k_x - \cos k_y)$. For the above parameters we obtain
$\Delta_d = 0.67t$. With respect to real materials this value is
exaggerated; however, our choice of parameters was motivated by the desire to have
the large mobility gap $\gamma_0$ (which according to Lee\cite{lee} scales with 
$\Delta_d^{1/2}$), necessary to study localization numerically in a finite system.

In Fig.\ \ref{ds} we display the amplitudes of $d$- and extended $s$-wave
order parameters defined as\cite{soininen}
\begin{eqnarray*}
d({\bf r})&=&{1\over 4}[\Delta_{\hat x}({\bf r})+\Delta_{-\hat x}({\bf r})
		-\Delta_{\hat y}({\bf r})-\Delta_{-\hat y}({\bf r})] \\
s({\bf r})&=&{1\over 4}[\Delta_{\hat x}({\bf r})+\Delta_{-\hat x}({\bf r})
		+\Delta_{\hat y}({\bf r})+\Delta_{-\hat y}({\bf r})] 
\end{eqnarray*}
in the vicinity of a single impurity with $V^{imp}=100t$. The $d$-wave order
parameter is suppressed at the impurity site and recovers its bulk value
over 2-3 lattice spacings, just as one would expect for a pair-breaking
impurity. The $s$-wave, which vanishes in the bulk, is nucleated near the
impurity site and displays an interesting real space structure with a 
``$d$-wave'' symmetry: it vanishes along the $|x|=|y|$ diagonals and changes
sign upon $90^\circ$ rotation. This behavior can be easily understood from 
the corresponding Ginzburg-Landau theory \cite{unpub}. We note that 
being of the form $d\pm s$ this state 
does not break time reversal symmetry.

For a finite density of impurities
we probe for localization in two different ways. First, for a given system
size and disorder configuration we evalute a generalized 
inverse participation ratio\cite{thouless},
\begin{equation}
a_n={\langle |u_n|^4\rangle + \langle |v_n|^4\rangle
     \over (\langle |u_n|^2\rangle + \langle |v_n|^2\rangle)^2},
\label{part}
\end{equation}
where $\langle \dots \rangle$ stands for the sum over all sites.
As a function of increasing system size $L$ this quantity decreases as $\sim 1/L^2$
for extended states and approaches a constant value $\sim(a/\xi_L)^2$ for a 
state localized within the characteristic length $\xi_L$.
Second, we study the sensitivity of states to boundary conditions
by applying a uniform phase twist $\chi$ accross the length of the system. We 
consider wavefunctions with the ``twisted'' 
boundary condition along the $x$-direction, 
$\Psi(x+L,y)=e^{i\chi}\Psi(x,y)$, and periodic along $y$, and we compute
the stiffness\cite{thouless,hatsugai} of the $n$-th eigenvalue
\begin{equation}
\kappa_n={1\over 2} \left({\partial^2 E_n\over \partial \chi^2}\right)_{\chi=0}.
\label{stiff}
\end{equation}
Extended states will be sensitive to twist, and
thus $\kappa_n$ is expected to be large. Localized states with $\xi_L\ll L$
will be insensitive to twist and  $\kappa_n$ is expected to be small.

\begin{figure}
\epsfxsize=8.0cm
\epsffile{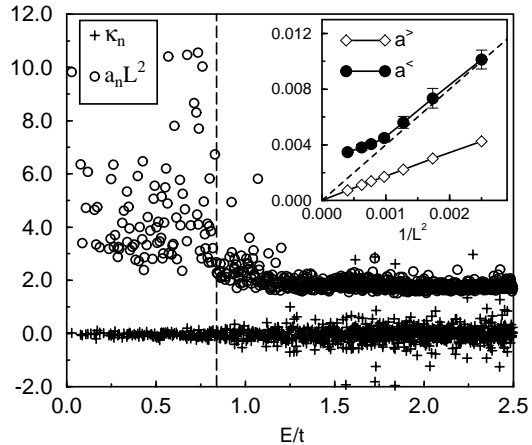}
\caption[]{ Inverse participation ratio $a_n$ and stiffness $\kappa_n$ 
plotted as a function of energy for $L=40$ and $n_i=0.06$. Dashed line marks the
mobility gap ${\gamma_0}=0.84$ estimated from the theory of Lee\cite{lee}.
Inset: Finite size scaling of the inverse participation ratio averaged over the 
energies below ($a^<$) and above ($a^>$) the mobility gap ${\gamma_0}$. 
 The error bars reflect the scatter of the 
data from 6 different impurity configurations.
Solid and dashed lines are guides to the eye only.}
\label{loc}
\end{figure}

We have studied systems with $n_i=0.015$ and $n_i=0.06$ 
of strong repulsive impurities with
$V^{imp}=100t$. In Fig.\ \ref{loc} we display typical results for the inverse
participation ratio $a_n$ and stiffness $\kappa_n$ as a function of energy for 
$n_i=0.06$.  
There is a pronounced qualitative difference in $a_n$ between the low energy 
states below $E/t \simeq 1$ and the high energy states. 
This is strongly suggestive  
of a mobility gap with a magnitude close to that predicted by Lee;\cite{lee} 
$\gamma_0\approx 0.84t$ for the parameters listed above.
While not so pronounced, this mobility gap can also be observed in the behavior 
of $\kappa_n$.
For lower density of impurities, $n_i=0.015$, we obtain similar results with 
the apparent mobility gap reduced by roughly a factor of two. This is in agreement
with Lee's prediction that $\gamma_0 \sim n_i^{1/2}$.

Even stronger evidence for the existence of the 
mobility gap can be obtained by analyzing the 
system size dependence of $a_n$. To this end we have carried out calculations for 
$L=20,24,28,32,36,40,50$ and six independent impurity configurations for each 
size. The inset of Fig.\ \ref{loc} shows two quantities, 
$a^<\equiv\langle\langle a_n \rangle\rangle_{E<\gamma_0}$ and 
$a^>\equiv\langle\langle a_n \rangle\rangle_{E>\gamma_0}$, defined as
averages of the inverse participation ratios $a_n$
taken over the indicated range of energy, as a function of $1/L^2$.
Data points for $a^>$ lie on a straight line which extrapolates to zero for 
$L\to\infty$, just as one would expect for the extended states. Values of
$a^<$ appear to extrapolate to a finite value of $a^<_\infty\approx 0.003$, 
which implies
that these states are localized with $\xi_L/a\approx 20$. We note that this 
value of $\xi_L$ is in a reasonable agreement with the rough
estimate given in Ref.\ [4], which gives $\xi_L/a\approx 38 $.

We have inspected visually the amplitudes of wavefunctions $u_n({\bf r})$ and
$v_n({\bf r})$ for signs of localization. We have found that, in agreement with  the
above analysis, these states are spatially localized below the mobility gap
with the characteristic length scale of about $20-30a$, 
and appear to be extended at higher energies.

The question naturally arises why we do not observe delocalization effects 
below the inverse mobility gap $\gamma_c$ predicted in Ref.\ \cite{balatsky}.
In order to clarify this issue we have studied  quasiparticle states associated
with a single impurity. For the model parameters described above ({\em i.e.,}
strong coupling) we found no evidence for the anisotropic cross-shaped states
predicted in Ref.\ \cite{balatsky95}. Thus, it is not surprising that for 
a finite density of impurities the delocalization mechanism based on 
long-ranged overlaps is inoperative and all we find is a sharp mobility edge
below which all the states are strongly localized. Furthermore we found that
in this regime it is very important to treat the order parameter 
self-consistently; a separate calculation with spatialy uniform ({\em i.e.},
not relaxed) order parameter yielded wavefunctions with different spatial
distributions and energy eigenvalues.

For the sake of completeness the above analysis was repeated for a different
set of model parameters which may be closer to the realistic values for cuprates
\cite{wang}: $V_0=1.05t$, $V_1=-1.05t$, $\mu=-0.36t$, which imply smaller
magnitude of the bulk gap, $\Delta_d\simeq 0.066t$. In this case we have indeed
found the predicteded highly anisotropic quasiparticle state, however only
the electron part of the wavefunction, $v_n({\bf r})$, exhbited the predicted
tails along the diagonal directions. The amplitude of the hole part, $u_n({\bf r})$,
was found to vanish along these diagonals with two tails running parallel to
it. Such a structure was recently also obtained from the $T$-matrix approximation
\cite{markku} and it is not clear how it affects the analysis of the 
localization problem in a system
with finite impurity concentration. Within our numerical method we were unable
to address this question since for the above parameters the estimated localization
length \cite{lee} would be much larger than the maximum system size we can 
handle.  Since in this case virtually identical
results were obtained with a spatially uniform order parameter, the method
employed previously by Xiang \cite{xiang1} may be suitable to address this
problem.

Thus, the evidence presently available allows for two distinct scenarios
with regards to the existence of the inverse mobility edge.  
Either ({\em i}) there is a crossover from the strong coupling regime
which has (as we have explicitly demonstrated) $\gamma_c=0$ to the weak coupling
regime with $\gamma_c >0$, or, ({\em ii}), it may be that $\gamma_c=0$ for all
coupling strengths which could be explained as a consequence of the complicated 
structure of the quasiparticle states discussed above
that was not included in the original analysis
of the many impurity problem \cite{balatsky}.

To conclude, we have presented clear evidence for strong localization by
impurities in $d$-wave superconductors at low energies. Compared to earlier
analytical work\cite{lee,balatsky} our approach 
treats impurity scattering exactly within BdG theory and accounts for order 
parameter relaxation near the impurity sites in a fully self-consistent way.
The present model can easily be extended to finite temperatures and 
frequencies, thus opening posibilities for the analysis of the temperature
dependent penetration depth, infrared conductivity, critical temperature,
and other physically interesting quantities in the presence of disorder.

The authors are grateful to M. I. Salkola for numerous helpful suggestions,
and to A. V. Balatsky, V. J. Emery, R. J. Gooding, J. M. Wheatley and T. Xiang 
for valuable discussions.
This work has been partially supported by the Natural Sciences and
Engineering Research Council of Canada and by the Ontario Centre for Materials 
Research.

\end{document}